\documentclass[aps,pre,preprint,groupedaddress,floatfix]{revtex4}
\usepackage{graphicx,amssymb}
\begin{document}

\title{Crystal nucleation along an entropic pathway: Teaching liquids how to transition.} 
\author{Caroline Desgranges and Jerome Delhommelle}
\affiliation{Department of Chemistry, University of North Dakota, Grand Forks ND 58202}
\date{\today}

\begin{abstract}
We combine machine learning (ML) with Monte Carlo (MC) simulations to study the crystal nucleation process. Using ML, we evaluate the canonical partition function of the system over the range of densities and temperatures spanned during crystallization. We achieve this on the example of the Lennard-Jones system by training an artificial neural network using, as a reference dataset, equations of state for the Helmholtz free energy for the liquid and solid phases. The accuracy of the ML predictions is tested over a wide range of thermodynamic conditions, and results are shown to provide an accurate estimate for the canonical partition function, when compared to the results from flat-histogram simulations. Then, the ML predictions are used to calculate the entropy of the system during MC simulations in the isothermal-isobaric ensemble. This approach is shown to yield results in very good agreement with the experimental data for both the liquid and solid phases of Argon. Finally, taking entropy as a reaction coordinate and using the umbrella sampling technique, we are able to determine the Gibbs free energy profile for the crystal nucleation process. In particular, we obtain a free energy barrier in very good agreement with the results from previous simulation studies. The approach developed here can be readily extended to molecular systems and complex fluids, and is especially promising for the study of entropy-driven processes.
\end{abstract}

\maketitle

\section{Introduction}

Entropy is central to many phenomena in chemistry, physics, biology and materials science. For instance, entropy is a key concept in our understanding of molecular association processes, such as, e.g., the dimerization of insulin~\cite{tidor1994contribution}, of pattern formation on nanoparticle surfaces~\cite{singh2007entropy}, of the partitionning behavior of small molecules in lipid bilayers~\cite{maccallum2006computer}, and of self-assembly processes~\cite{whitelam2015hierarchical,escobedo2017optimizing}. Furthermore, entropy provides a measure of the amount of coded information contained in the canonical genetic code~\cite{nemzer2017shannon}, as well as a thermodynamic characterization of supercooled liquids~\cite{angell1997entropy,starr2003prediction} and amorphous materials~\cite{berthier2017configurational,yan2018entropy}. Entropic effects are also crucial for mixture separation during adsorption processes~\cite{schenk2001separation,krishna2002entropy}, for the design of protein geometries to facilitate the active transport of ions~\cite{rubi2017entropy} as well as in protein-protein binding for signal transduction and molecular recognition.~\cite{sun2017interaction,caro2017entropy}. 

Since entropy provides a measure of the amount of order, or conversely, of disorder, within a system~\cite{frenkel2014order}, it is often used to characterize the level of organization of this system. As such, entropy-driven processes have been the focus of intense research in recent years, and have led to many insightful discoveries. Such phenomena include the entropy-driven self-assembly in charged lock-key particles~\cite{odriozola2016entropy}, entropy-driven segregation processes of polymer-grafted nanoparticles~\cite{zhang2017entropy}, the entropy-driven crystallization of DNA grafted nanoparticles~\cite{thaner2015entropy}, polymers~\cite{karayiannis2009entropy}, proteins~\cite{derewenda2006entropy}, as well as of atomic~\cite{piaggi2017enhancing} and molecular crystals~\cite{gobbo2018nucleation}. Furthermore, such studies have shed light on the interplay between competing processes, e.g. entropy-driven segregation and crystallization~\cite{zha2016entropy,desgranges2014unraveling}, that can lead to complex molecular mechanisms, such as two-step nucleation processes for silicon~\cite{desgranges2011role} and clathrate colloidal crystals~\cite{lee2018entropy}. Recent work has also highlighted the key role played by entropy during the nucleation of liquid droplets~\cite{desgranges2016free1,desgranges2016free2}, the formation of capillary bridges during condensation processes in nanotubes~\cite{vishnyakov2003nucleation,remsing2015pathways,desgranges2017free} and during the nucleation of cavitation bubbles~\cite{menzl2016effect}.

A full characterization of these entropic effects hinges on the determination of entropy along the transition pathway~\cite{meirovitch2007recent}. However, such calculations are not straightforward, since a direct determination of the entropy $S=-k_B \sum_i P^B_i \ln P^B_i$ requires knowing the Boltzmann probability $P^B_i$, in which $i$ denotes a system configuration~\cite{karplus1981method,schafer2000absolute,shell2008relative,foley2015impact}. Alternatively, free energy calculations can be carried out to determine the entropy change along a given pathway, using either thermodynamic integration~\cite{lu2001accuracy,white2004simulation,cheluvaraja2004simulation,preisler2016configurational} or flat histogram methods~\cite{singh2003surface,Pond,desgranges2012eval1,desgranges2012eval2}. Other possible routes involve entropy expansions in terms of the correlation functions~\cite{baranyai1989direct,jakse2016excess} or the analysis of the dynamical response of the system, through a Fourier transform of the velocity autocorrelation function~\cite{lin2003two,lin2010two}. In this work, we adopt a different approach and take advantage of machine learning techniques to determine on-the-fly the entropy of the system as it undergoes crystal nucleation. Machine learning is now widely used in applications as diverse as modeling molecular atomization energies~\cite{rupp2012fast}, in density functional theory~\cite{bartok2013machine} and molecular docking~\cite{ballester2010machine}. In particular, artificial neural networks (ANNs) have been shown to give excellent results for the free energy landscape of a wide range of systems, including peptides, biomolecules and polymers~\cite{guo2018adaptive,sidky2018learning,mansbach2015machine}. Here, we use ML to predict the canonical partition function of the system over a wide range of temperature and densities spanned during the crystallization process. To achieve this, we train ANNs to model accurately the Helmholtz free energy of the Lennard-Jones system, as given by an equation of state obtained by Johnson {\it et al.} for the liquid phase~\cite{Johnson}, and by the equation of state of van der Hoef~\cite{vanderhoef} for the solid phase. Then, using the predictions provided by the ANN for Helmholtz free energy, we are able to determine the entropy $S$ of the system. The advantages of this approach are three-fold, since it allows us to (i) determine the canonical partition function, and thus the Helmholtz free energy as a function of only two variables, density and temperature, along the crystallization process, (ii) calculate the entropy of a system over a wide range of conditions, and (iii) define entropy as a reaction coordinate to follow the onset of order in the system, here during a crystal nucleation event. To assess the first point, we show that the ML predictions are accurate over a wide range of conditions and yields estimates for the canonical partition function in excellent agreement with the results from flat-histogram simulations. Second, we compare the entropy obtained on the Lennard-Jones system to experimental data available for Argon~\cite{Vargaftik}, and show that the entropy obtained using our method is in very good agreement with the data on both liquid and solid Argon. Third, we simulate the crystal nucleation process in the isothermal-isobaric ensemble using entropy as a reaction coordinate. Specifically, we implement the umbrella sampling technique within NPT Monte Carlo simulations, and compare our results with prior work in the field~\cite{tenWolde} using geometric parameters, such as the bond orientational order parameter $Q_6$ defined by Steinhardt {\it et al.}~\cite{Steinhardt}.

The paper is organized as follows. In the next section, we present the approach developed in this work, starting with the type of machine learning technique used here, and how entropy is evaluated and can be used as a reaction coordinate in Monte Carlo simulations. We then assess the reliability and accuracy of the method by testing the ability of this approach to model the Helmholtz free energy over a wide range of thermodynamic conditions. We also show how the ML predictions yield an accurate picture of the thermodynamics of the system. In particular, we compare the ML canonical partition function of the system to the results from prior Wang-Landau simulations. Then, we determine the entropy of the system and assess the accuracy of the method through a comparison to the available experimental data for both solid and liquid phases. We also study the crystal nucleation process from a supercooled liquid of Lennard-Jones particles, determine the free energy barrier of nucleation along the entropic pathway and compare our results to those from previous work. We finally draw the main conclusions from this work in the last section.

\section{Simulation Method}

\subsection{Machine learning technique and entropy calculations}

\begin{figure}
\begin{center}
\includegraphics*[width=10cm]{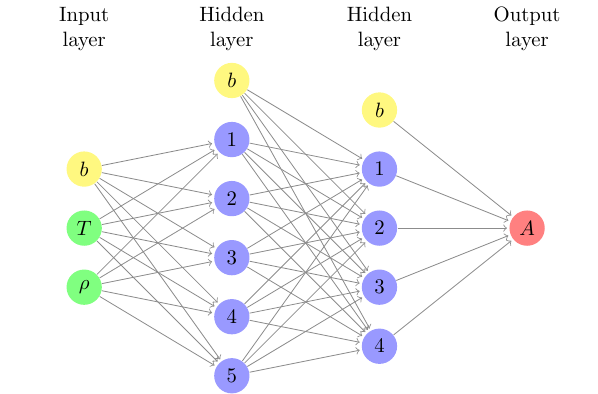}
\end{center}
\caption{Structure of the ANN used in this work for the determination of the Helmholtz free energy $A$ for a given temperature $T$ and density $\rho$.}
\label{Fig1}
\end{figure}

The first step in our approach consists of using machine learning (ML) to model accurately the canonical partition function $Q(N,V,T)$, or equivalently the Helmholtz free energy $A=-k_BT \ln Q(N,V,T)$, for densities and temperatures spanned during crystallization. For this purpose, we train an artificial neural network using as reference data the Helmholtz free energy of the Lennard-Jones system from the Johnson, Zollweg and Gubbins equation of state ($JZG-EOS$) for the fluid phase~\cite{Johnson} and from the van der Hoef equation of state ($VDH-EOS$) for the face-centered cubic crystal~\cite{vanderhoef}. The structure of the ANN used in this work is given in Fig.~\ref{Fig1}. 

The ANN we employ includes 4 layers, the input and output layers, as well as two hidden layers. The input layer is composed of 2 input neurons, corresponding to the two input parameters, temperature $T$ and density $\rho$, for which we need to determine the Helmholtz free energy of the system. The next two layers are hidden layers, for which we have used, respectively, $5$ and $4$ neurons. We add that we include the usual bias node $b$ for the input layer and every hidden layer~\cite{desgranges2018new,guo2018adaptive,sidky2018learning,mansbach2015machine}. Finally, the output layer only contains a single neuron, since we aim here to predict the Helmholtz free energy for any $(T,\rho)$ set. The machine-learned Helmholtz free energy $A^{ML}$ can then be obtained through the following equation~\cite{behler2016perspective,geiger2013neural}
\begin{equation}
A^{ML}= f_4 [b_3 +\sum_{l=1}^{4} W(3,4,l,1) f_3 (b_2 +\sum_{j=1}^{5} W(2,3,j,l) f_2 [(b_1 + W(1,2,1,j)T+W(1,2,2,j) \rho])]
\label{AML}
\end{equation}
in which $W(i-1,i,l,j)$ is the weight matrix connecting neuron $l$ from layer $i-1$ to neuron $j$ from layer $i$, the $f_i$ functions are the activation functions, chosen as the $tanh$ function for $i=1,2,3$ and the linear function for $i=4$, The weights for the ANN are trained during an iterative process, going first through a forward pass and then a backpropagation algorithm. Eq.~\ref{AML} therefore provides a way to estimate the Helmholtz free energy not only for the equilibrium phases, i.e. the solid and liquid which serve to train the ANN, but also for the intermediate, nonequilibrium, states~\cite{koss2017free} that connect the starting point, the metastable liquid, to the top of the free energy barrier, corresponding to the formation of the critical nucleus.

The entropy of the system $S$ can be calculated from the canonical partition $Q(N,V,T)$ function through the following statistical mechanical expression
\begin{equation}
S= {U + k_B T \ln Q (N,V,T) \over T }
\label{entropyQ}
\end{equation}
in which $U$ is the internal energy of the system.

Eq.~\ref{entropyQ} suggests a route for the determination of entropy during the simulations, in a form that can be readily used as a reaction coordinate for the nucleation process. For this purpose, starting from Eq.~\ref{entropyQ}, we define $S_i$ for each configuration $i$ generated during the $(N,P,T)$ simulations as
\begin{equation}
S_i= {U_i - A^{ML}(\rho_i,T) \over T }
\label{entropy}
\end{equation}
in which $U_i={U^{pot}_i + {3 \over 2} N k_B T}$ is the sum of the interaction (potential) energy and of the ideal gas (kinetic) contribution to the internal energy and $A^{ML}(\rho_i,T)$ are the ML predictions for the Helmholtz free energy at a density $\rho_i={N \over V_i}$ and a temperature $T$. We assess in Section~III the accuracy and reliability of Eq.~\ref{entropy} by testing it against experimental data for Argon, for which the Lennard-Jones model performs well, and against the results from flat-histogram simulations.

\subsection{Exploring an entropic pathway}

The next step consists of using the entropy, determined from our machine learning approach, as a reaction coordinate. Previous work has shown that the umbrella sampling ($US$) technique~\cite{Torrie} is a very successful approach to study the onset of nucleation, both for the vapor-liquid  transition~\cite{ten1999numerical,mcgrath2010vapor,desgranges2016free1,desgranges2018calculating,xi2016sparse} and the solid-liquid transition~\cite{tenWolde,auer2001prediction,radhakrishnan2003nucleation,blaak2004crystal,desgranges2006molecular,desgranges2006insights,desgranges2007controlling,desgranges2007molecular,punnathanam2006crystal,sanz2007evidence,yi2009molecular,filion2010crystal,desgranges2014unraveling,desgranges2018unusual,ketzetzi2018crystal,prestipino2018barrier}. We therefore build on our previous work~\cite{desgranges2016free1,desgranges2016free2,desgranges2017free}, in which we simulated the nucleation of liquid droplets using $US$ in the grand-canonical $\mu VT$ simulations. Here, since we focus on crystal nucleation, the very low acceptance rates for the insertions/deletions of atoms preclude the use of $\mu VT$ simulations. We therefore implement the $US$ technique, using the machine-learned entropy as a reaction coordinate, within isothermal-isobaric $NPT$ Monte-Carlo simulations. The US method is a non-Boltzmann sampling technique~\cite{Torrie,Allen} that consists in applying a US potential, usually chosen as a harmonic function of a reaction coordinate, to sample configurations with a very low Boltzmann weight. Once the  simulation data have been collected, the US potential is then subsequently removed to obtain the free energy profile for the process, specifically here the Gibbs free energy profile since this approach is implemented in the isothermal-isobaric ensemble~\cite{Torrie,tenWolde}. The ANN weights allow for the calculation of entropy for a given temperature $T$ and density $\rho$, making the use of the machine-learned entropy very well suited to serve as a reaction coordinate for processes occurring under isothermal-isobaric conditions, such as crystal nucleation. From a practical standpoint, we start from a metastable supercooled liquid (parent phase) and carry out simulations for gradually decreasing values of entropy to promote the formation of the critical nucleus. In each of these simulations, we use an US potential with the following functional form
\begin{equation}  
U_{bias}={1 \over 2} k (S_i-S_0)^2
\end{equation}
in which $S_0$ is the target value for the entropy, $S_i$ is provided using Eq.~\ref{entropy} for a configuration $i$ of the system and $k$ is a spring constant. Then, we collect the simulation results under the form of an histogram, that yields the probability associated with each value of the entropy once the US potential has been removed~\cite{Torrie,Allen}. Repeating the simulations for steadily decreasing target values for the entropy allows us to build the Gibbs free energy profile for the nucleation process as discussed in prior work~\cite{tenWolde,Auer,desgranges2018unusual}.

\subsection{Technical details}

The first stage of this work involves training the ANN. This is carried out using a learning rate set to $0.2$, as well as a training dataset that includes $11,970$ data points, generated by the $VDH-EOS$ and the $JZG-EOS$. For the solid ($VDH-EOS$), we include data for densities ranging from $0.94$ to $1.2$ and temperatures from $0.1$ to $2$ (here, the density and temperature are given in units reduced with respect to the Lennard-Jones parameters $\epsilon$ and $\sigma$). For the fluid phase ($JZG-EOS$), we include data for temperatures ranging from $0.7$ to $6$ . The ANN training is deemed to be complete when the error estimate becomes less than $5 \times 10^{-7}$. Once the ANN has been trained, we perform two types of simulations. First, we test the accuracy of the ML predictions for the Helmholtz free energy, the canonical partition function and for the entropy on simulations carried out on systems of $500$ Lennard-Jones particles for the fluid and solid phases. Second, we carry out simulations of the crystal nucleation process using the $US$ technique under isothermal-isobaric conditions and with the entropy as a reaction coordinate and refer to these simulations as $NPT-S$ simulations in the rest of this work.  Simulations of crystal nucleation are carried out on system of $3,000$ Lennard-Jones particles for a reduced temperature $T^*=0.86$ and a reduced pressure $P^*=5.68$, which correspond to a supercooling of 22\%~\cite{vanderhoef}. All simulations are carried out within a $NPT$ Monte Carlo (MC) framework. Two types of random moves are implemented, corresponding to the random translation of a single atom (99\% of the attempted MC moves) or a random change in the volume of the cubic simulation cell, leading to a rescaling of the positions of all atoms (1\% of the attempted MC moves). The maximum displacement and volume change are set such that 50\% of the attempted MC moves are accepted. The usual periodic boundary conditions are applied~\cite{Allen}, and tail corrections beyond a cutoff radius of $3\sigma$ are employed.

\section{Results and Discussion}

\subsection{ML predictions for the Helmholtz free energy and the canonical partition function.}
\begin{figure}
\begin{center}
\includegraphics*[width=9cm]{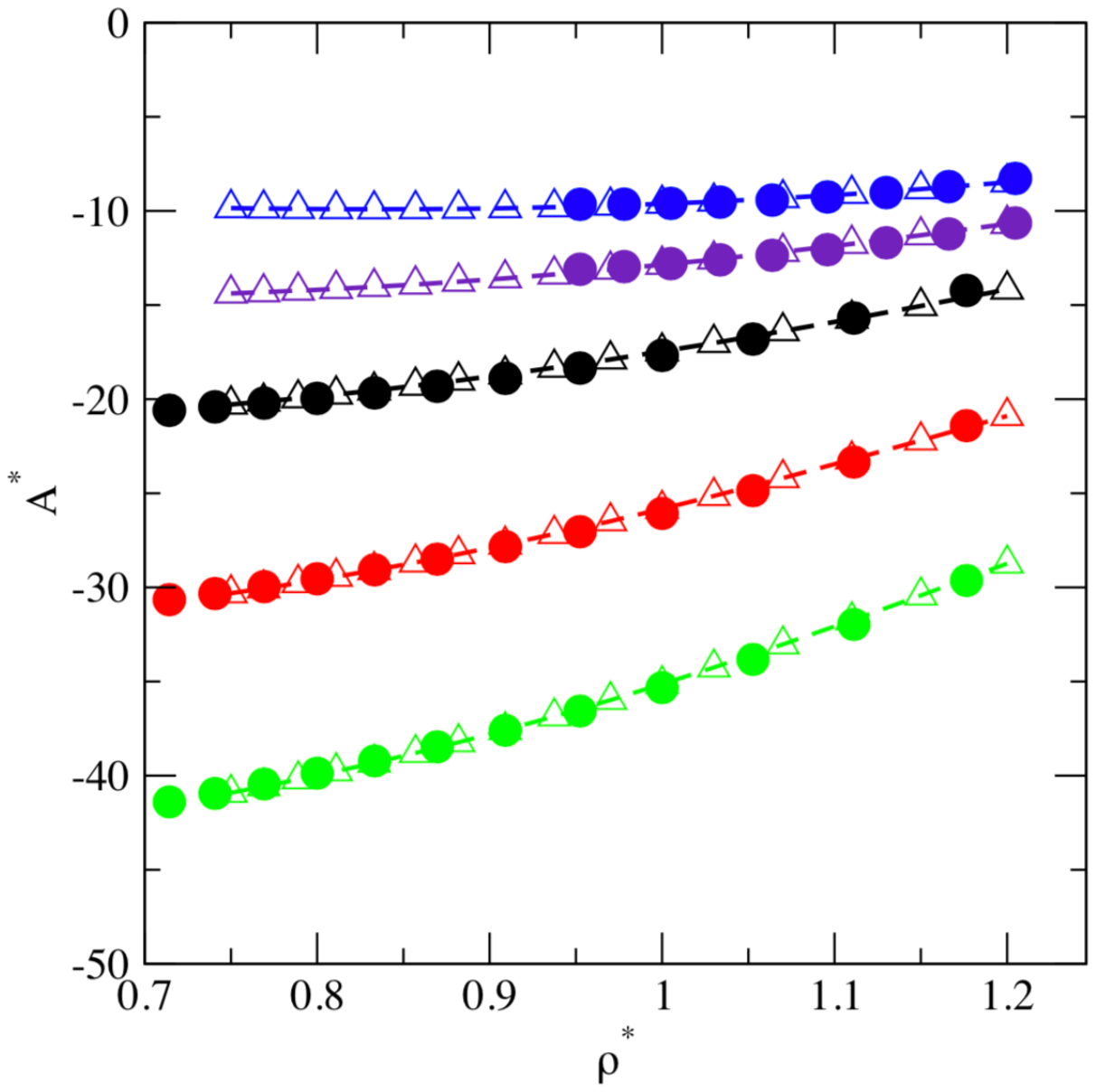}
\end{center}
\caption{Helmholtz free energy as a function of the reduced density. Results from the $JZG-EOS$ are shown as filled circles for a reduced temperature of $T^*=2$ (in black), $T^*=3$ (in red) and $T^*=4$ (in green) for the liquid, while results from the $VDH-EOS$ are plotted for reduced temperatures of $T^*=0.75$ (in blue) and $T^*=1.35$ (in purple). The ML predictions for the corresponding conditions are shown as open triangles.}
\label{Fig2}
\end{figure}

We start by presenting the results obtained for the ML predictions of the Helmholtz free energy. Fig.~\ref{Fig2} shows a plot of the reduced Helmholtz free energy $A^*$ against the reduced density $\rho^*$. In this graph, we compare the ML predictions, obtained using to Eq.~\ref{AML}, to the data from the two equations of state we use for the Lennard-Jones system, namely the $JZG-EOS$ for the liquid and the $VDH-EOS$ for the solid. Fig.~\ref{Fig2} shows that there is an excellent agreement between the ML predictions and the EOS data across a wide temperature range, from $T^*=0.75$ to $T^*=4$. Furthermore, we see that the weights of the trained ANN allow us to obtain very accurate results for both the liquid phase and the solid phase (FCC crystal). More specifically, focusing e.g. on the agreement between the $VDH-EOS$ and ML predictions for $T^*=1.35$, we find a root mean square deviation (RMSD) of $0.11$. Turning to the agreement between the $JZG-EOS$ and the ML predictions for the liquid at $T^*=3$, we obtain a RMSD of $0.15$. Both examples show that ML predictions for the Helmholtz free energy are in excellent agreement with the EOS data.

\begin{figure}
\begin{center}
\includegraphics*[width=10cm]{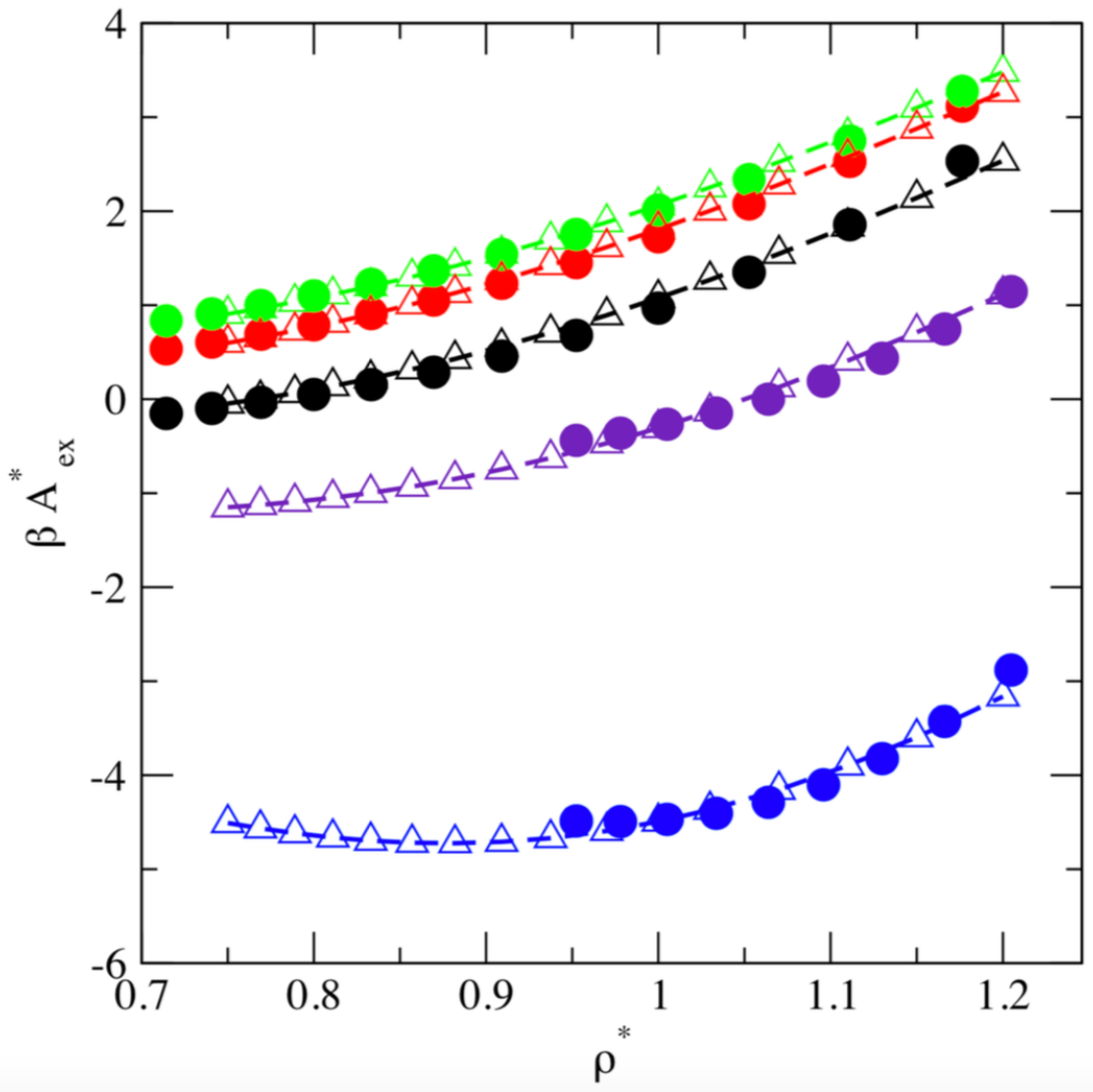}
\end{center}
\caption{Excess Helmholtz free energy against density for the Lennard-Jones system. Same legend as in Fig.~\ref{Fig2}.}
\label{Fig3}
\end{figure}

Second, we analyze how the ML predictions perform for the excess Helmholtz free energy, which is also provided by the EOS data. We show in Fig.~\ref{Fig3} the variations of the product of the inverse temperature by the excess Helmholtz free energy ($A^{*}_{ex}$) for both phases, and for reduced temperatures ranging from $T^*=0.75$ to $T^*=4$. The quantity $\beta A^{*}_{ex}$ behaves in a dramatically different way when compared to the reduced Helmholtz free energy. The largest values for $\beta A^{*}_{ex}$ are now obtained for the highest temperature ($T^*=4$) and the lowest $\beta A^{*}_{ex}$ values are reached for $T^*=0.75$. This is the expected behavior, since the attractive interactions between LJ particles are predominant as the temperature of the system decreases, and the solid becomes the stable phase. Furthermore, this effect becomes more pronounced as a result of the scaling by the inverse temperature $\beta$ at low temperatures. This qualitative trend is well captured by both the EOS data and the ML predictions. From a quantitative standpoint, in line with the results obtained for $A^*$, we find that the ML predictions for the excess Helmholtz free energy are in very good agreement with the $JZG-EOS$ and $VDH-EOS$ data. Considering the same state points as for $A^*$, we find a RMSD of $0.05$ between the EOS data and the ML predictions at $T^*=3$ for the liquid, and of $0.08$ at $T^*=1.35$ for the solid. This confirms the ability of the trained ANN to calculate accurately the Helmholtz free energy of both the liquid and solid phases for the Lennard-Jones system.

\begin{figure}
\begin{center}
\includegraphics*[width=10cm]{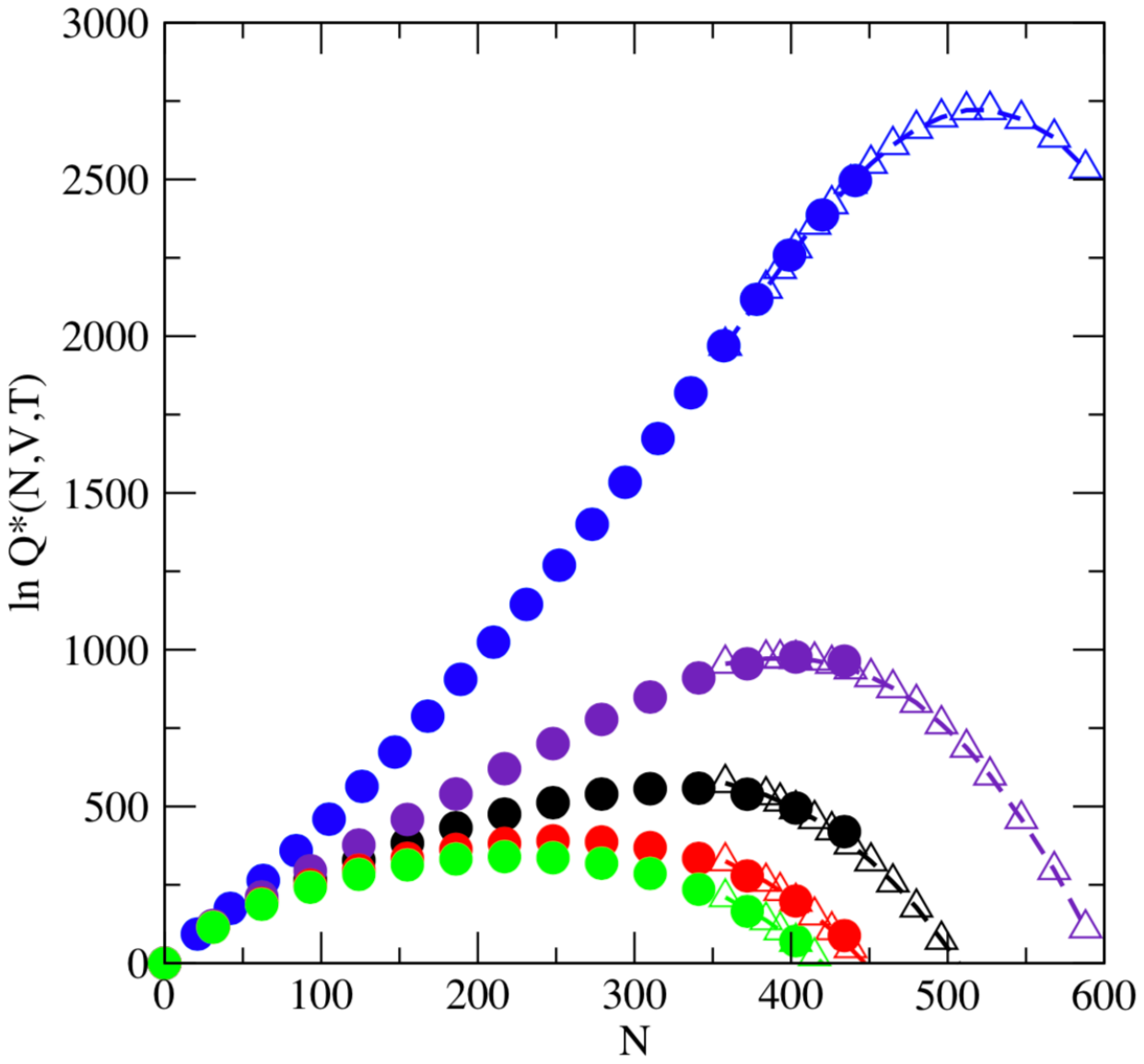}
\end{center}
\caption{Logarithm of the reduced canonical partition functions $Q^*(N,V,T)$ against $N$ , the number of LJ particles. Filled circles are the results from prior work using EWL simulations~\cite{desgranges2012eval1}, for reduced temperatures $T^*=0.75$ (in blue), $T^*=1.35$ (in purple), $T^*=2$ (in black), $T^*=3$ (in red) and $T^*=4$ (in green). Open triangles are the ML predictions.}
\label{Fig4}
\end{figure}

To assess further the performance of the ML predictions, we now compare their ability to predict the value of the canonical partition function. The partition function can be obtained from the Helmholtz free energy from the following equation 
\begin{equation}
A=-k_B T \ln Q(N,V,T)
\end{equation}
with 
\begin{equation}
Q(N,V,T)= { {V^N} \over {N! \Lambda^{3N}}} \int \exp (-\beta U( \mathbf{\Gamma})) d\mathbf{\Gamma}
\end{equation}
where $\Lambda$ is the de Broglie wavelength and the integration is performed over the coordinates of the $N$ atoms of the system ($\mathbf{\Gamma}$ denoting a specific configuration of the system).
This relation therefore allows us to determine $Q^{ML}(N,V,T)$ from the ML predictions for the Helmholtz free energy. We present in Fig.~\ref{Fig4} a comparison between the ML derived partition function $Q^{ML}(N,V,T)$ and the partition function obtained in prior simulation work using the Expanded Wang-Landau (EWL) simulation method~\cite{desgranges2012eval1}. Fig.~\ref{Fig4} shows that the ML predictions and the EWL simulation results are in very good agreement over the whole temperature range from $T^*=0.75$ to $T^*=4$, as demonstrated for densities corresponding to the liquid phase in the plot. Interestingly, the ML predictions also provide an insight into how the partition function varies as a function of $N$ for very large densities. This range of densities corresponding to the solid phase is not accessible through EWL simulations. This is because EWL simulations are implemented within the grand-canonical ensemble and, as such, require the insertion and deletion of atoms during the simulation, which becomes very difficult at high densities.

\subsection{Entropy calculations for liquid and solid phases.}

\begin{figure}
\begin{center}
\includegraphics*[width=10cm]{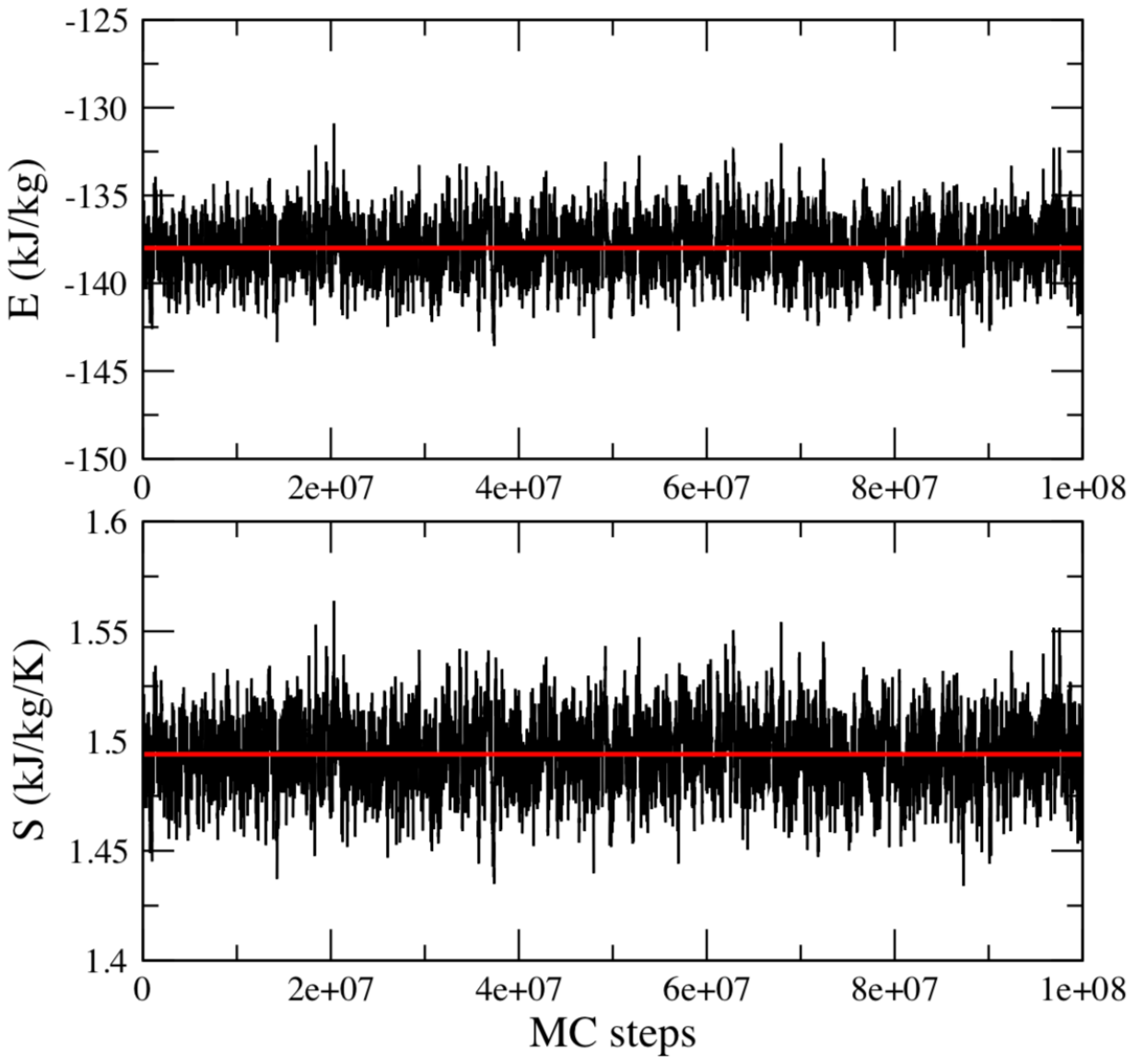}
\end{center}
\caption{Monte Carlo NPT simulations of liquid Argon at $T=95$~K and $P=5$~bar. Variation of the interaction energy (top) and of entropy (bottom) as a function of the number of MC steps. In both plots, the red line indicated the average value.}
\label{Fig5}
\end{figure}

Now that the reliability of the ML predictions for the Helmholtz free energy has been examined, we now look at how these perform for the determination of the entropy of the system. For this purpose, we carry out the first type of simulation discussed in the Methods section. In order to compare to experimental for the liquid~\cite{Vargaftik} and for the solid~\cite{rabinovich1988thermophysical}, we present results  in real units, using the conventional Lennard-Jones parameters for Argon ($\sigma=3.4$~\AA, $\epsilon=117.05$~K and $m=40$~g/mol). We show in Fig.~\ref{Fig5} the results of NPT simulations carried out at $T=95$~K and $P=5$~bar for the liquid. In this graph, we plot the variation of the interaction energy during the simulation, as well as the entropy calculated through Eq.~\ref{entropy}. These plots exhibit the behavior expected, once the simulation has converged, and provide an estimate for the extent of the fluctuations around the average value during the simulation. 

How do the simulation results compare to the experimental data? To assess this in detail, we carry out a series of NPT simulations of the liquid phase for several state points, with pressure ranging from $P=5$~bar to $P=100$~bar and temperature ranging from $T=90$~K to $T=100$~K. The results are provided in Table~\ref{tab1}. Liquid densities predicted by the simulations are found to be in good agreement with the experimental data (within 0.03 $g/cm^3$), which confirms that the NPT simulations have converged, since the Lennard-Jones (LJ) potential is known to be a good model for Argon under these conditions. Looking closely at the thermodynamic function enthalpy (H), we also see that there is a good agreement between the experimental data and the simulation results with the maximum deviation of $0.6$~kJ/kg occurring at the the lowest temperature ($T=90$~K). This also provides an estimate of the accuracy with which the LJ potential models the thermodynamic properties of real Argon. Turning now to the entropy calculations, we find that there is a very good agreement between the experimental data and our calculations, since the maximum deviation is also observed at $T=90$~K in line with the findings for H and is of $0.02$~kJ/kg/K. Moving on to the Gibbs free energy (G), we obtain a good agreement between the experimental data and the simulation results, with deviations of less than 1\% across the entire range of conditions studied.

 \begin{table}[hbpt]
 \caption{Liquid Argon: comparison between the experimental data~\cite{Vargaftik} and the simulation results (this work) for various state points. Temperatures are given in $K$, $\rho$ in $g/cm^3$, S in $kJ/kg/K$, H and G in $kJ/kg$. Standard deviations are of the order of $3 \times 10^{-3}$~g/cm$^{-3}$ for the density, of $10^{-2}$~kJ/kg/K for $S$, and of $0.1$~kJ/kg for $H$ and $G$.}
 \begin{tabular}{ccccccccc}
 \hline
 $ $ & $$ & $$ & $$ & $$ & $P=5~bar$ & $$ & $$ & $$  \\
 \hline
  $T$ &$ \rho_{exp} $ &$ \rho_{sim} $ & $S_{exp}$ & $S_{sim} $& $H_{exp}$ & $H_{sim}$ & $G_{exp}$ & $G_{sim}$\\
 \hline
   90 & 1.377 & 1.375 & 1.420 & 1.442 & -113.2 & -113.8 & -241.0 & -243.5 \\
   95 & 1.344 & 1.342 & 1.477 & 1.494 & -108.0 & -108.0 & -248.3 & -249.9 \\
 100 & 1.311 & 1.308 & 1.532 & 1.543 & -102.6 & -102.3 & -255.8 & -256.6 \\
\hline
\hline
 $ $ & $$ & $$ & $$ & $$ & $P=100~bar$ & $$ & $$ & $$  \\
 \hline
   90 & 1.405 & 1.403 & 1.391 & 1.414 & -109.1 & -109.6 & -234.3 & -236.8  \\
   95 & 1.376 & 1.374 & 1.446 & 1.462 & -104.0 & -104.0 & -241.4 & -243.0 \\
 100 & 1.347 & 1.344 & 1.498 & 1.509 & -98.9 & -98.6 & -248.7 & -249.6 \\
 \hline
 \end{tabular}
 \label{tab1}
 \end{table}

\begin{figure}
\begin{center}
\includegraphics*[width=10cm]{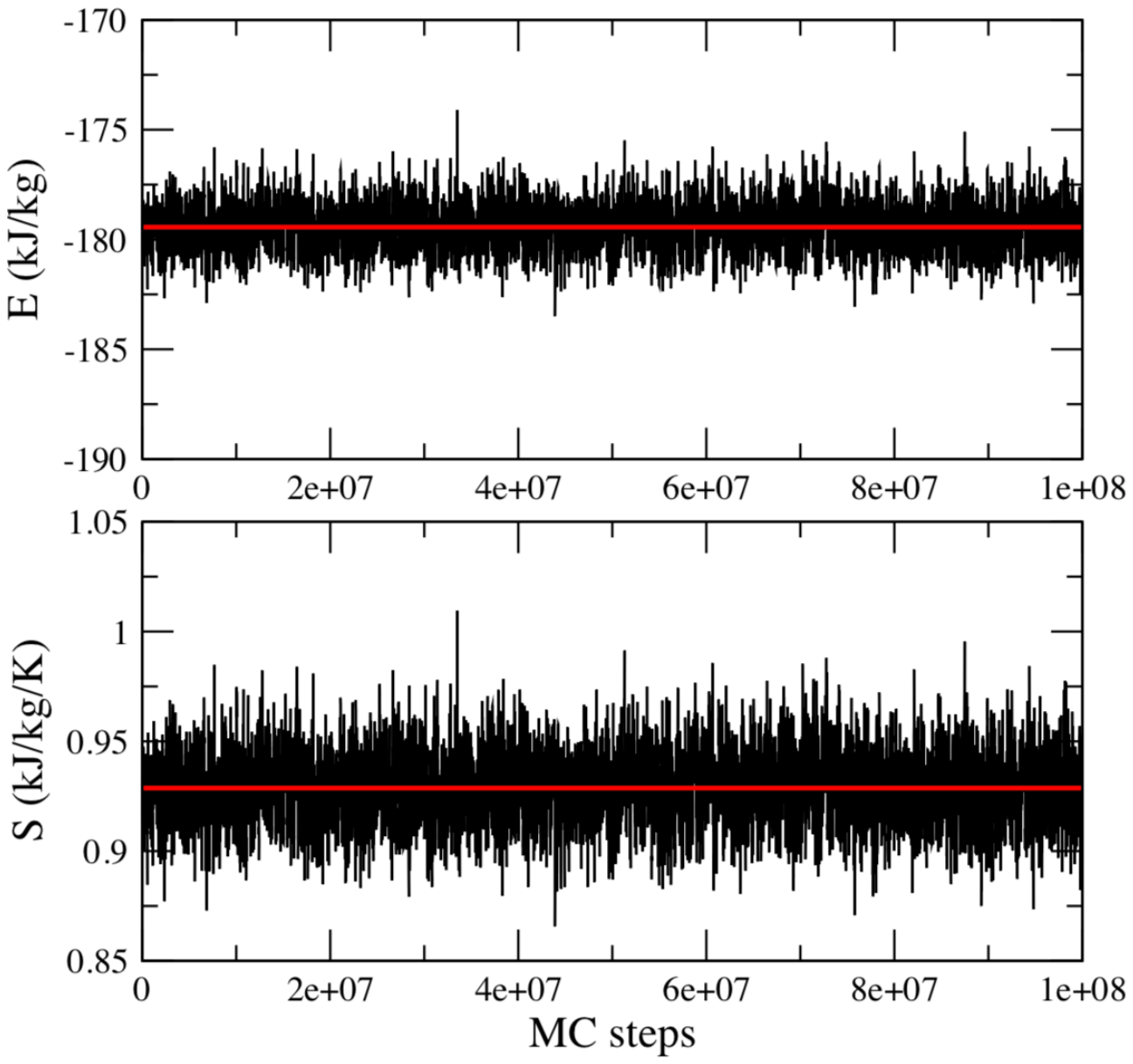}
\end{center}
\caption{Monte Carlo NPT simulations of solid Argon at $T=80$~K and $P=100$~bar. Same legend as in Fig.~\ref{Fig5}.}
\label{Fig6}
\end{figure}

We now analyze the results obtained for the solid phase. First, we show in Fig.~\ref{Fig6} the variation of the interaction energy during a NPT simulation of a FCC crystal at $T=80$~K and $P=100$~bar. We observe a behavior similar to that of the liquid (shown in Fig.~\ref{Fig5}), but with a lower interaction energy as a result of the increased attractive interactions that takes place in the solid phase. Turning to the variations of entropy during the simulation, we also find that the entropy fluctuates around its average value. As expected, we find this average value to be lower than for the liquid, as a result of the greater amount of order found in the crystal.
  
As for the liquid, we perform a series of NPT simulations on the FCC phase for temperatures ranging from $T=70$~K to $T=80$~K and for pressures from $P=1$~bar to $P=100$~bar. The simulation results are given in Table~\ref{tab2}, together with the experimental data for the solid. The results for the densities show that there is an overall good agreement between the experimental data and the simulation results for the density, with deviations of less than $0.02$~$g/cm^3$ for the state points studied here. In line with results for the liquid, we also obtain a good agreement for the thermodynamic functions, with the simulation results for the enthalpy exhibiting a deviation of up to 1\% for the enthalpy, of up to 1.1\% for the entropy and up to 1\% for the Gibbs free energy. This confirms that the entropy calculations carried out using the ML predictions for the Helmholtz free energy, together with Eq.~\ref{entropy}, provide accurate results for both the liquid and the solid phases, as established through this comparison with the experimental data.
 
\begin{table}[hbpt]
 \caption{Solid Argon: comparison between the experimental data~\cite{rabinovich1988thermophysical} and the simulation results (this work) for various state points. Temperatures are given in $K$, $\rho$ in $g/cm^3$, S in $kJ/kg/K$, H and G in $kJ/kg$ (standard deviations are of the same order as for the liquid).}
 \begin{tabular}{ccccccccc}
 \hline
 $ $ & $$ & $$ & $$ & $$ & $P=1~bar$ & $$ & $$ & $$  \\
 \hline
  $T$ &$ \rho_{exp} $ &$ \rho_{sim} $ & $S_{exp}$ & $S_{sim} $& $H_{exp}$ & $H_{sim}$ & $G_{exp}$ & $G_{sim}$\\
 \hline
   70 & 1.664 &1.681 & 0.825 & 0.823 & -160.4 & -161.9 & -218.2 & -219.5 \\
   75 & 1.649 & 1.662 & 0.874 & 0.886 &-156.5 & -157.6 & -222.1 & -224.1 \\
   80 & 1.633 & 1.642 & 0.932 & 0.950 & -152.4 & -153.0 & -227.0 & -229.0 \\
\hline
\hline
 $ $ & $$ & $$ & $$ & $$ & $P=100~bar$ & $$ & $$ & $$  \\
 \hline
   70 & 1.675 &1.693 & 0.815 & 0.806 & -155.2 & -156.9 & -212.2 & -213.3  \\
   75 & 1.661 & 1.676 & 0.867 & 0.867 & -151.4 & -152.7 & -216.4 & -217.4 \\
   80 & 1.645 & 1.656 & 0.919 & 0.929 & -147.4 & -148.3 & -220.9 & -222.6 \\
 \hline
 \end{tabular}
 \label{tab2}
 \end{table}
 
 \subsection{Crystal nucleation using entropy as reaction coordinate.}

\begin{figure}
\begin{center}
\includegraphics*[width=10cm]{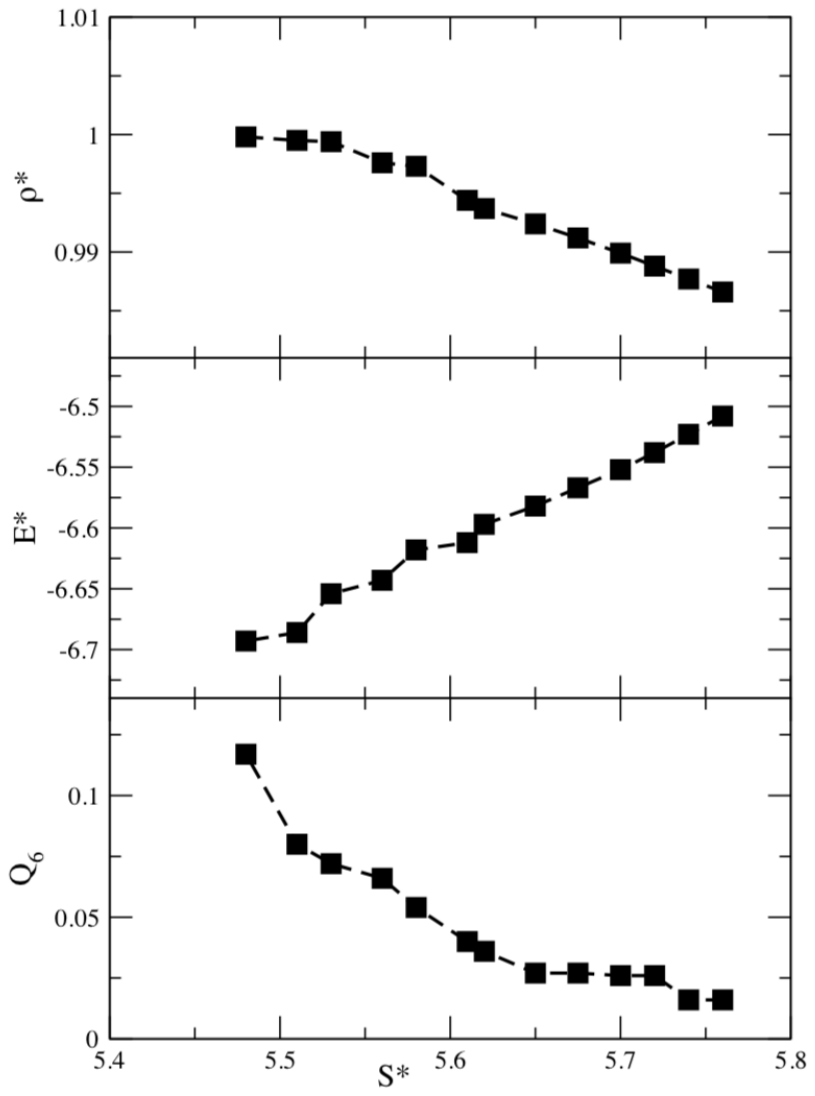}
\end{center}
\caption{Variation of (a) the reduced density, (b) the reduced interaction energy and (c) $Q_6$ as a function of the reduced entropy for the successive NPT-S simulations.}
\label{Fig7}
\end{figure}

We finally turn to the last type of simulations carried out in this work, and perform NPT-S simulations of crystal nucleation in the supercooled liquid of Lennard-Jones particles at $T^*=0.86$ and $P^*=5.68$. We start from the supercooled liquid, for which we obtain an average density of $\rho^*=0.989$. Using the ANN trained in the first part of this work, together with Eq.~\ref{AML}, we have a Helmholtz free energy $A^*=-10.31$, which yields an entropy $S^*=5.71$ through Eq.~\ref{entropy}. From there, we start carrying out NPT-S simulations, through umbrella sampling, to study the onset of crystal nucleation. We perform $13$ successive simulations with overlapping windows, each window being associated with a decreasing value for the target entropy. We plot in Fig.~\ref{Fig7} the variation of the density during the successive NPT-S simulations, with each data point on the graph corresponding to the average density for a simulation window. We find that, as entropy gradually decreases, the density of the system steadily increases, with an overall increase amounting to 1.3\% over the entire range of entropies sampled. We also show in Fig.~\ref{Fig7} how the interaction energy is impacted by the decrease in entropy. We observe that the interaction energy steadily decreases with entropy, resulting in a overall change of -2.8\% over the whole process. Furthermore, to characterize the order within the system, we calculate the value taken by the Steinhardt order parameter~\cite{Steinhardt} $Q_6$ over each window. We plot the evolution of $Q_6$ as the function of the decreasing entropy in Fig.~\ref{Fig7} and find that $Q_6$ increases as entropy decreases, indicating that crystalline order starts to develop within the system. Overall, the increase in density and $Q_6$, combined with the decrease in interaction energy, point to an increased level of organization within the system. This is consistent with the decrease of the entropy of the system, and with the onset of crystal nucleation.

\begin{figure}
\begin{center}
\includegraphics*[width=9cm]{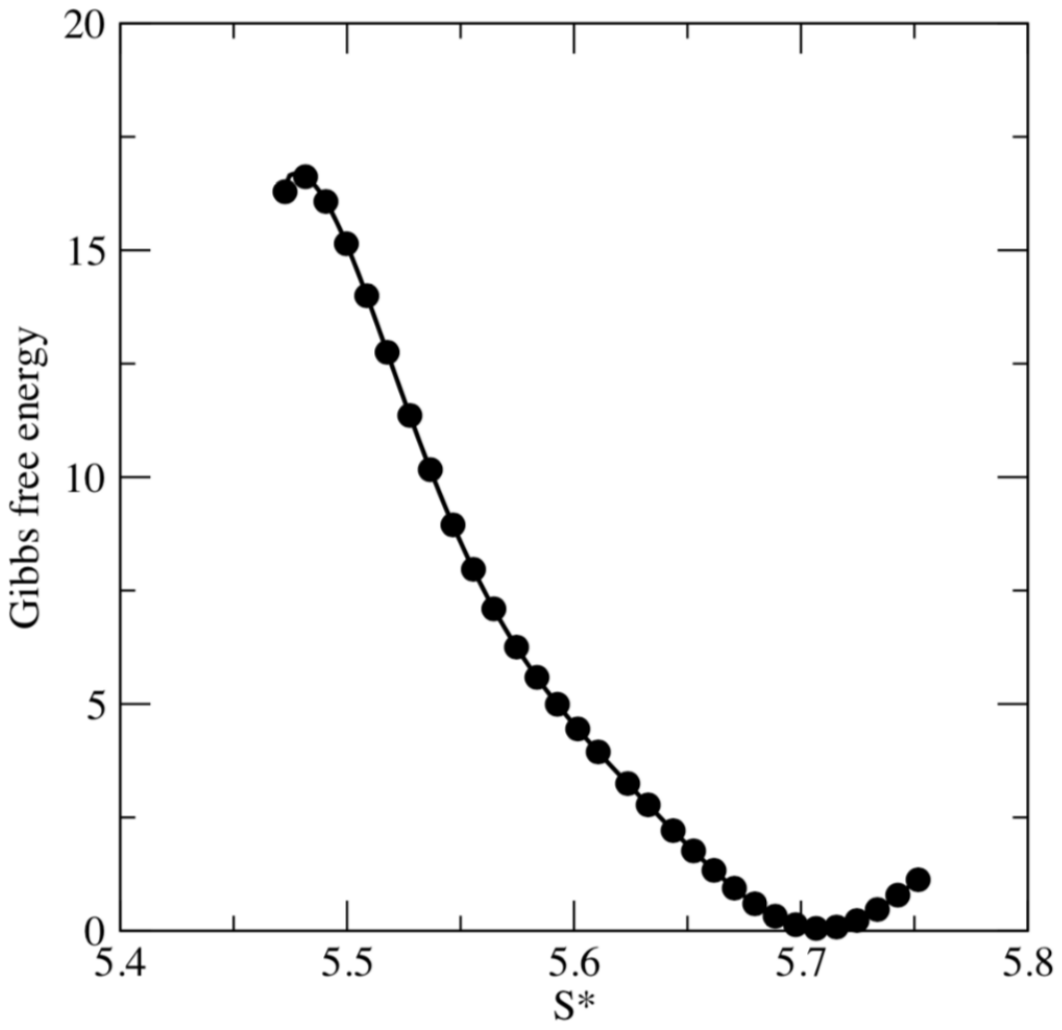}(a)
\includegraphics*[width=7cm]{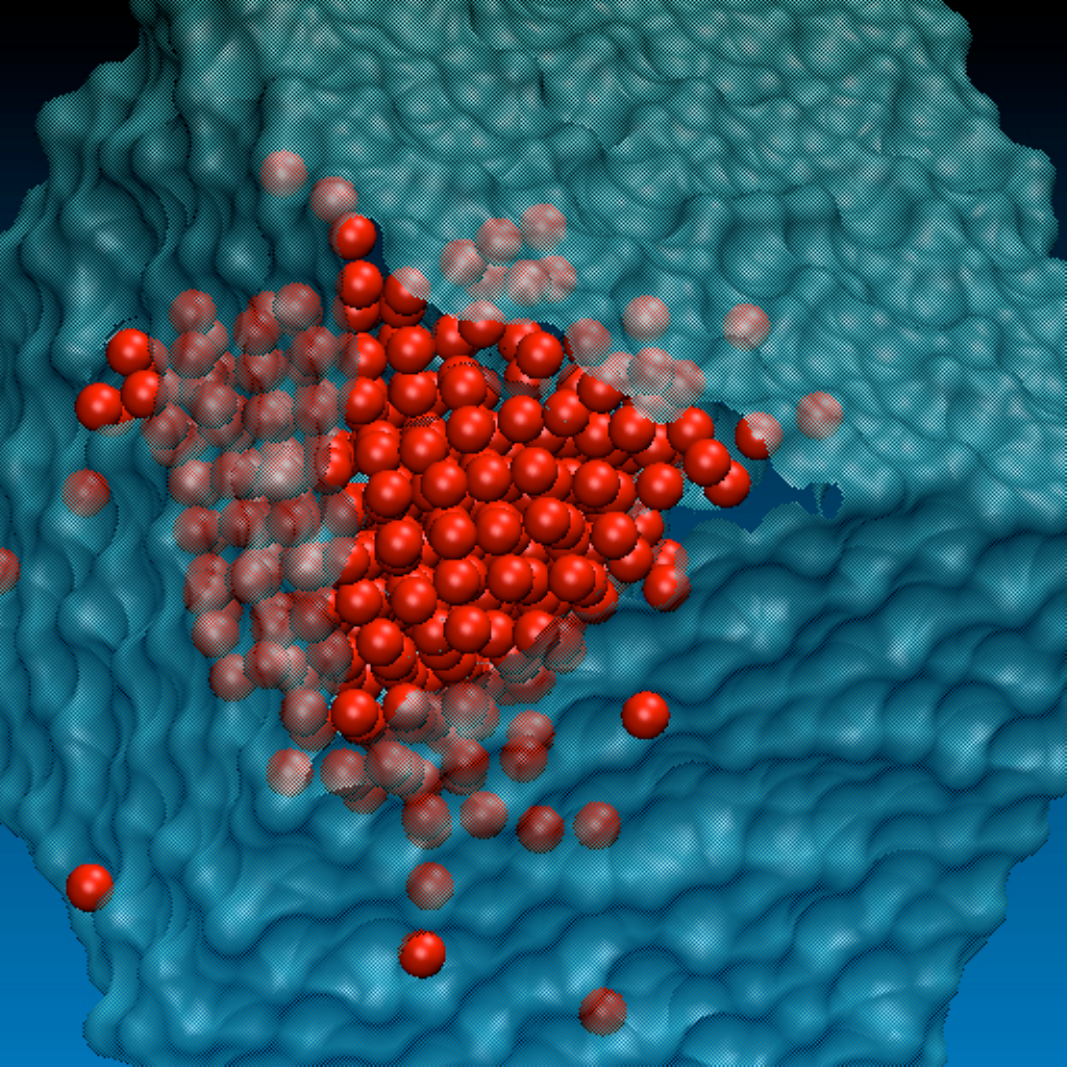}(b)
\end{center}
\caption{(a) Crystal nucleation of the Lennard-Jones system at $T^*=0.86$ and $P^*=5.68$: Gibbs free energy of nucleation as a function of the reduced entropy $S^*$. (b) Snapshot of a crystal nucleus (in red) of a critical size formed within a supercooled liquid (in cyan) of Lennard-Jones particles.}
\label{Fig8}
\end{figure}

Using the umbrella sampling results, we calculate the free energy profile for the entire process and present it in Fig.~\ref{Fig8}(a). The free energy profile shows that the starting point for the nucleation process is the metastable supercooled liquid with $S^*=5.71$, which is associated with a local minimum in free energy as shown in Fig.~\ref{Fig8}(a). Then, as entropy decreases, the organization within the system starts to develop. During this stage, we observe that the free energy increases, until the top of the free energy barrier is reached for $S^*=5.5$. At this point, a critical nucleus has formed. We show in Fig.~\ref{Fig8}(b) a snapshot of the critical nucleus, surrounded by the liquid (solid-like particles are identified using the same criteria as in previous work). Looking now at the height of the free energy barrier obtained for this process, we find that it is of $17 \pm 2$ ~$k_BT$. This is consistent with the results from prior simulations on the Lennard-Jones system~\cite{tenWolde}, which estimates a barrier of roughly $20$~$k_BT$ for the supercooling considered in this work. These results show that our approach, which relies on the use of entropy as a reaction coordinate, can indeed be used to simulate the onset of crystal nucleation within a supercooled liquid.

\section{Conclusions}
In this work, we use machine learning to determine the Helmholtz free energy of a system and, in turn, to calculate its entropy using molecular simulation. This is achieved by training an artificial neural network using, as a training dataset, previously published equations of state for the Lennard-Jones system for both the fluid and solid phases. Once the ANN has been trained, we are able to determine the Helmholtz free energy as a function of density and temperature. Our approach is validated by testing the ML predictions over a wide range of thermodynamic conditions. In particular, these predictions for the Helmholtz free energy lead us to calculate the canonical partition function, thereby providing a direct comparison with results from recent flat-histogram simulations on the Lennard-Jones system. The second stage of this work consists of using these ML predictions to calculate the entropy of the system. This is validated by performing MC simulations under NPT conditions. The simulation results are found to be in very good agreement with experimental data on liquid and solid Argon, a system that is very well modeled by the Lennard-Jones potential. Finally, we use entropy as a reaction coordinate in umbrella sampling simulations, and show that a gradual decrease in entropy results in the formation of a crystal nucleus within a supercooled liquid. The analysis of the crystal nucleation process along the entropic pathway is found to be consistent with the results from previous simulation studies. This approach can be readily extended to nucleation processes for molecular systems and complex fluids, and is also especially promising for the study of entropy-driven organization processes.   

\bibliography{NPT_S}

\end{document}